%% file: D0R32.tex
\documentclass[aps,prl,twocolumn,showpacs,superscriptaddress,groupedaddress]{revtex4}  % for review and submission
\usepackage{graphicx}  % needed for figures
\usepackage{dcolumn}   % needed for some tables
\usepackage{bm}        % for math
\usepackage{amssymb}   % for math

\newcommand{\ptmax}{p_{T\text{max}}}
\newcommand{\ptmin}{p_{T\text{min}}}

\newcommand{\as}{\alpha_s}

\newcommand{\pythia}{{\sc pythia}}
\newcommand{\sherpa}{{\sc sherpa}}

\newcommand{\Rcone}{R_{\text{cone}}}
\newcommand{\Rtt}{R_{\text{3/2}}}

\newcommand{\ppbar}{p{\bar{p}}}

\begin{document}

% the following line is for submission, including submission to the arXiv!!
\hspace{5.2in} \mbox{Fermilab-Pub-12-500-E}

\title{Measurement of the ratio of three-jet to two-jet cross sections \\
       in {\boldmath${p\bar{p}}$} collisions
  at {\boldmath$\sqrt{s}=1.96$}~TeV}
\input author_list.tex       % D0 authors (remove the first 3 lines
                             % of this file prior to submission, they
                             % contain a time stamp for the authorlist)
                             % (includes institutions and visitors)
\date{September 5, 2012}

\begin{abstract}
We present a measurement of the ratio
of multijet cross sections
in $\ppbar$ collisions at $\sqrt{s}=1.96\,$TeV
at the Fermilab Tevatron Collider.
The measurement is based on a data set corresponding to an
integrated luminosity of $0.7\,$fb$^{-1}$
collected with the D0 detector.
The ratio of the inclusive three-jet to two-jet cross sections, $\Rtt$,
has been measured as a function of the jet transverse momenta.
The data are compared to QCD predictions in different approximations.
Popular tunes of the \pythia\ event generator do not agree with the data, while \sherpa\ provides a reasonable description
of the data.
A perturbative QCD prediction in next-to-leading order in the strong coupling
constant, corrected for non-perturbative effects, gives a good description
of the data.\end{abstract}

\maketitle

In hadron-hadron collisions, production rates of collimated sprays of hadrons,
called jets, with large transverse momenta with respect to the beam axis ($p_T$) are sensitive to both
the dynamics of the fundamental interaction and to the partonic structure of the initial-state hadrons.
The latter is usually parameterized in parton distribution functions (PDFs) of the hadrons.
We observed large sensitivities to PDFs in our measurement of the differential
three-jet cross section as a function of the three-jet invariant mass~\cite{zdanek}.
Studies dedicated to the dynamics of the interaction are therefore preferably based on quantities
which are minimally sensitive to the PDFs. Such quantities can be constructed as ratios of cross sections, for which the
sensitivity to the PDFs is reduced.
One class of such quantities is the ratio of multijet cross sections.
The two-jet~\cite{dijetnlo} ,  three-jet~\cite{threejetnlo}, and four-jet~\cite{fourjetnlo} cross sections have been computed
in perturbative Quantum Chromodynamics (pQCD) up to next-to-leading order (NLO) in the strong coupling constant $\as$.
The ratio of the inclusive three-jet cross section to the inclusive two-jet cross section, $\Rtt$, provides a test of the corresponding NLO pQCD predictions.
Previous measurements of $\Rtt$\, in processes with initial state hadrons, have been made in $ep$ collisions at the
HERA Collider at DESY~\cite{desyr32}, and in hadron-hadron collisions at the SPS Collider at CERN~\cite{spsr32}, the Fermilab Tevatron Collider~\cite{d0run1}, and at the Large Hadron Collider at CERN~\cite{cmsr32,atlasmulti}.

This letter presents the first measurement of $\Rtt$\ in $p\bar{p}$ collisions at a center-of-mass
energy of $\sqrt{s}=1.96$\,TeV. The results are presented as a function of the highest jet $p_T$ in the event, $\ptmax$,
for four minimum values of the second highest and (for three-jet events) third highest jet $p_T$, $\ptmin$.
The data sample, collected with the D0 detector during 2004--2005
in Run~II of the Fermilab Tevatron Collider, corresponds to
an integrated luminosity of $0.7\,$fb$^{-1}$.

Jets are defined by the Run~II midpoint cone jet algorithm~\cite{run2cone}
with a cone radius of $\Rcone = 0.7$.
Rapidity is related to the polar scattering angle $\theta$
with respect to the beam axis
by $y=0.5 \ln \left[ (1+\beta \cos \theta) / (1-\beta \cos \theta) \right]$,
with $\beta=|\vec{p}| / E$.
The inclusive $n$-jet event sample (for $n=2,3$) is defined
by all events in which the $n$ highest $p_T$ jets have $p_T > \ptmin$ and  $|y|<2.4$.
The separations in the plane of rapidity and azimuthal angle $\phi$,
$R_{\sl jj}=\sqrt{(\Delta y)^2 + (\Delta \phi)^2}$
between the $n$ highest $p_T$ jets are required to be
larger than twice the cone radius
($R_{\sl jj} > 2\Rcone$).
The rapidity requirement restricts the jet phase space to the region
where jets are well-reconstructed in the D0 detector
and the energy calibration is known to
1.2--2.5\% for jets with $50<p_T<500\,$GeV.
The separation requirement strongly reduces the phase space
for which the $n$ highest $p_T$ jets had overlapping cones which were
split during the overlap treatment
of the jet algorithm.

The ratio of inclusive three-jet to two-jet cross sections,
\[\Rtt(\ptmax,\ptmin) = {d\sigma_{\text{3-jet}}(\ptmin)/d\ptmax\over{d\sigma_{\text{2-jet}}(\ptmin)/d\ptmax}},\]
is less sensitive to experimental and theoretical
uncertainties than the individual cross sections due to cancellations of correlated uncertainties.
Here $\Rtt(\ptmax,\ptmin)$ is measured
for $\ptmin$ requirements of 30, 50, 70, and 90\,GeV
in bins of $\ptmax$,
in the interval $80\,{\rm GeV} < \ptmax < 500\,{\rm GeV}$,
with the additional requirement of $\ptmax > \ptmin + 30$\,GeV.
This additional requirement
ensures that there is sufficient phase space for the
second and third jet so that corrections due to the
experimental $p_T$ resolution remain small.
Given the definitions above for the inclusive $n$-jet event samples,
at each $\ptmax$\ value the inclusive three-jet event sample is a
subset of the inclusive two-jet event sample.
Therefore $\Rtt(\ptmax,\ptmin)$ represents the conditional probability
for a two-jet event (at $\ptmax$) to contain a third jet with $p_T > \ptmin$.

% *********************************************************************
% ***************   Measurement
% *********************************************************************

A detailed description of the D0 detector can be found in
Ref.~\cite{d0det}.
The event selection, jet reconstruction, jet energy and
momentum correction in this measurement follow closely
those used in our recent measurements of inclusive jet, two-jet, and three-jet  distributions~\cite{2008hua,2012jet,2009mh,mandy,markusrdr,zdanek}.
The primary tool for jet detection is the
finely segmented uranium-liquid argon calorimeter that
has almost complete solid angle coverage
$1.7^\circ \lesssim \theta \lesssim 178.3^\circ$~\cite{d0det}.
Events are triggered by a single high $p_T$ jet above a particular
threshold.
In each $\ptmax$ bin, events are taken from a single trigger
which is chosen such that the trigger efficiency
is above 99\% for two-jet and for three-jet events.
Using triggers with different prescale values results
in integrated luminosities of
1.54\,pb$^{-1}$ for $\ptmax<120\,$GeV,
17\,pb$^{-1}$ for $120<\ptmax<140\,$GeV,
73\,pb$^{-1}$ for $140<\ptmax<175\,$GeV,
0.5\,fb$^{-1}$ for $175<\ptmax<220\,$GeV,
and 0.7\,fb$^{-1}$ for $\ptmax>220\,$GeV.

% ********************************************************************
% **************   Jet Correction: JES, vertex, MET, jet ID
% ********************************************************************

The position of the $p\bar{p}$ interaction, reconstructed using a
tracking system consisting of silicon microstrip detectors~\cite{d0smt} and
scintillating fiber tracker located inside a $2\,\text{T}$ solenoidal
magnet~\cite{d0det},
is required to be within $50$\,cm of the detector center
along the beam direction.
The jet four-momenta are corrected for the response of the calorimeter,
the net energy flow through the jet cone,
energy from event pile-up and multiple $p\bar{p}$ interactions,
and for systematic shifts in rapidity due to detector effects~\cite{2012jet}.
Cosmic ray backgrounds are suppressed by requirements on
the missing transverse momentum in an event~\cite{2012jet}.
Requirements on characteristics of the shower shape are
used to suppress the remaining background due to electrons, photons,
and detector noise that mimic jets.
The efficiency for these requirements is above $97.5\%$,
and the fraction of background events is below $0.1$\% at all $\ptmax$.

The jet four-momenta reconstructed from calorimeter energy depositions
are then corrected, on average, for the response of the calorimeter,
the net energy flow through the jet cone,
additional energy from previous beam crossings, and
multiple $p\bar{p}$ interactions in the same event, but not for
muons and neutrinos~\cite{2008hua,2012jet}.
The absolute energy calibration is determined from
$Z \rightarrow e^+e^-$ events and the
$p_T$ imbalance in $\gamma$ + jet events in the region $|y| < 0.4$.
The extension to larger rapidities is derived from dijet events
using a similar data-driven method.
In addition, corrections in the range 2--4\% are applied that take
into account the difference in calorimeter response due to the
difference in the fractional contributions of quark and
gluon-initiated jets in the dijet and the $\gamma$ + jet event samples.
These corrections are determined using jets simulated
with the \pythia\ event generator~\cite{pythia} that have been
passed through a {\sc geant}-based detector simulation~\cite{geant}.
The total corrections of the jet four-momenta vary between 50\% and
20\% for jet $p_T$ between 50 and 400\,GeV.
An additional correction is applied for systematic shifts in $|y|$
due to detector effects~\cite{2008hua,2012jet}.
These corrections adjust the reconstructed jet energy to the
energy of the stable particles that enter the
calorimeter except for muons and neutrinos.

% ---------------------------------------------------------
% -----------     correction, simulation
% ---------------------------------------------------------
The $\Rtt$ distributions are corrected for instrumental effects
using a simulation of the D0 detector response based on
parameterizations of resolution effects in $p_T$, the polar and
azimuthal angles of jets, and jet reconstruction efficiencies.
The parameterizations are determined either from data or
from a detailed simulation of the D0 detector using {\sc geant}.
The parameterized simulation uses events generated with
\sherpa\~v1.1.3~\cite{Gleisberg:2008ta}
(including the tree-level matrix elements for two-jet, three-jet, and four-jet production)
using default settings and MSTW2008LO PDFs~\cite{Martin:2009iq},
and a sample of events, generated with \pythia\~v6.419 using tune QW~\cite{Albrow:2006rt} and CTEQ6.6 PDFs~\cite{CTEQ:2008}.
The events are subjected to the detector simulation
and are reweighted such that their simulated distributions
describe the differential two-jet and three-jet cross sections
in the $p_T$ and rapidity of each of the three highest $p_T$ jets in the
data.
To minimize migrations between $\ptmax$ bins due to resolution effects,
we use the simulation to obtain a rescaling function in $\ptmax$
that optimizes the correlation between the reconstructed and true values.
The rescaling function is applied to data and simulation.
The bin sizes in $\ptmax$ are chosen to be much
larger than the $p_T$ resolution.
The bin purity after $\ptmax$ rescaling, defined as the fraction
of all reconstructed events that are generated in the same bin,
is above $50\%$ for the two-jet and above $45\%$
for the three-jet event samples.
Bin efficiencies, defined as the fraction of all generated events
that are reconstructed in the same bin, are above
55\% for the two-jet and above 45\% for the three-jet event samples.

We use the simulation to determine correction factors for
the differential two-jet and three-jet cross sections in all $\ptmax$ bins,
taking the average of {\sc sherpa} and \pythia.
These include corrections for all instrumental effects, including
the energies of unreconstructed muons and neutrinos inside the jets.
The total correction factors for the differential cross sections are
between 0.92 and 1.0 for the two-jet and in the range 0.98--1.1
for the three-jet event samples.
The correction factors for the ratio $\Rtt$ are in the range
$0.9$--$1.2$.
Over most of the range, the corrections from the two models agree
within 3\%.
We take half the difference as an estimate of the model dependence of
the correction, taking into account the correlations between the uncertainties for the two sets of correction factors.
The corrected data are presented at the ``particle level''(jets formed from stable particles after fragmentation) as
defined in Ref.~\cite{Buttar:2008jx}.

\begin{figure*}[hbt]
    \begin{center}
   \includegraphics[width=15.6cm]{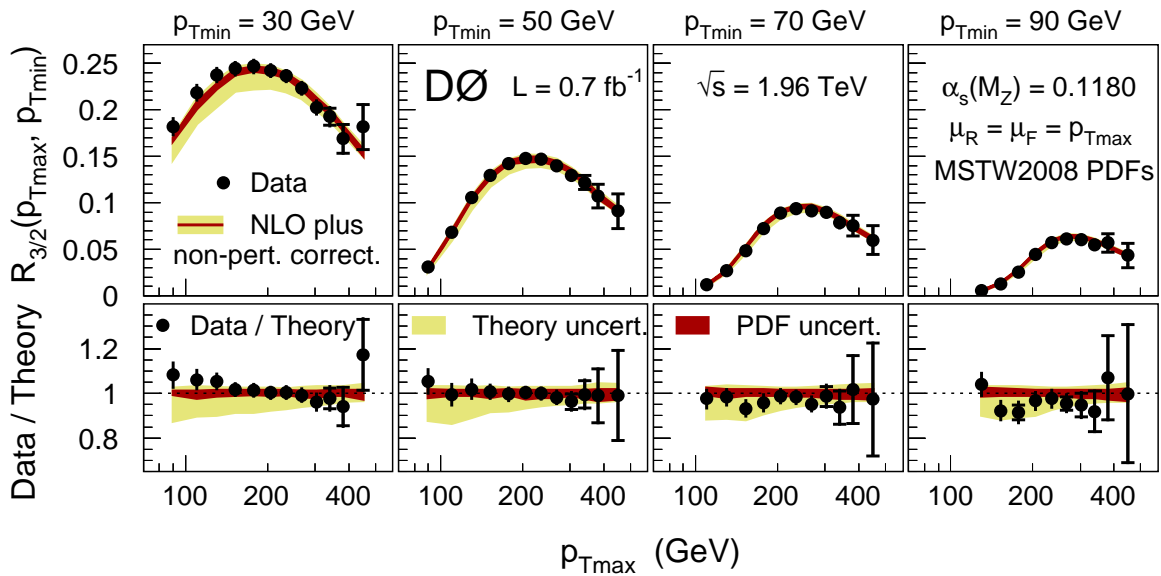}
  \caption{\label{fig:result}
  The measured $\Rtt$ results, compared to the predictions from NLO pQCD corrected for non-perturbative effects (top),
  and the ratio of data to theoretical predictions (bottom).
  The results are presented as
  a function of the highest jet $p_T$, $\ptmax$, for different $\ptmin$ requirements.
  The inner uncertainty bars represent the statistical uncertainties
  while the total uncertainty bars represent the quadratic sums of statistical
  and systematic uncertainties.
  (N.B. the inner uncertainty bars are within the markers for most data points.)
  }
  \end{center}
\end{figure*}

% ---------------------------------------------------------
% -----------     Comparison with theory
% ---------------------------------------------------------

The $\Rtt$\ measurement results are listed in
Table~\ref{tab:consistency} and Ref.~\cite{SuppMat},
and are displayed in Fig.~\ref{fig:result}
as a function of $\ptmax$ for different
$\ptmin$ requirements.
The ratio $\Rtt$ increases with increasing $\ptmax$ up to a maximum
value and decreases for higher $\ptmax$ values.
The position and the height of the maximum depend on the $\ptmin$ requirement
(for the $\ptmin$ choices in this analysis, the maximum appears at
$\ptmax$ values in the range 200--300\,GeV).
For a given value of $\ptmax$, three-jet final states have on average larger invariant masses than two-jet final states.
Therefore the three-jet cross section approaches the kinematic limit at lower $\ptmax$ than the two-jet cross section, resulting in the
decrease of $\Rtt$\ at large $\ptmax$. The initial increase of $\Rtt$\ with $\ptmax$\ reflects the increasing phase space for three-jet final states,
for a given $\ptmin$\ requirement. For higher $\ptmin$ requirements, the initial increase of $\Rtt$ occurs at higher $\ptmax$ values,
thereby shifting the position of the maximum.

\input{R32table.tex}

Theoretical calculations for $\Rtt$\ are computed as the product of NLO pQCD results and correction factors for non-perturbative
effects. Predictions of NLO pQCD are obtained from {\sc nlojet++}~\cite{NLOJET} using {\sc fastnlo}~\cite{FASTNLO}.
Jets are reconstructed using the FastJet~\cite{FASTJET} implementation of the D0 Run~II midpoint cone jet algorithm.
We use the two-loop approximation of the renormalization group equation for five quark flavors with $\as(M_Z) = 0.1180$ which is
close to the world average value of 0.1184~\cite{PDG}.
Results are computed using the MSTW2008NLO~\cite{mstw2008nlo}, the CT10~\cite{CT10}, and the NNPDF2.1~\cite{NNPDF} PDF sets.
For consistency, we always use those PDFs which have been obtained for $\as(M_Z) = 0.1180$.
The renormalization and factorization scales $\mu_R$\ and $\mu_F$\ are set to $\mu_0 = \ptmax$.
The scale uncertainties are computed by varying $\mu_R$ and $\mu_F$ independently
between $\mu_0/2$ and $2 \mu_0$ with the restriction that
$0.5 < \mu_R/\mu_F<2.0$. The uncertainties of the pQCD predictions due to the scale dependence
are between $-15$\% and $+5$\%.

The non-perturbative correction factors are the products of hadronization and underlying event corrections.
Both are estimated using \pythia\ with tunes DW~\cite{Albrow:2006rt} and AMBT1~\cite{AMBT1}.
Tune DW
uses $Q^2$-ordered parton showers and an older underlying event model,
while AMBT1 uses $p_T$-ordered parton showers and a newer underlying event model.
For each of these tunes, three event samples have been generated: parton shower level without an underlying event,
particle level without an underlying event, and particle level with an underlying event.
The hadronization corrections are estimated as the ratio of $\Rtt$\ at the particle level and
at the parton level (from the partons at the end of the parton shower). Both are obtained without an underlying event.
The underlying event correction is the ratio of the particle level
results with and without an underlying event. We use the average of the corrections obtained with tunes DW and AMBT1 as the central choice,
and quote half the spread as the uncertainty. The total non-perturbative correction factors are in the range
of 0.96--0.99 with uncertainties of less than 1\%.

\begin{table}[h]
\centering
\caption{\label{tab:r32chi2}
The $\chi^2$ values between theory and data for different $\ptmin$ requirements,
for different choices of $\mu_R$, $\mu_F$.}
\begin{ruledtabular}
\begin{tabular}{ccrrr}
$p_{T{\rm min}}$  & Number of   &   \multicolumn{3}{c}{$\chi^2$ for  $\mu_R=\mu_F=$} \\
                  & data points & $p_{T{\rm max}} / 2 $ & $p_{T{\rm max}}$ & $2p_{T{\rm max}}$ \\
% & points & $\mu_0 /2  $ & $\mu_0$ & $2 \cdot \mu_0$ \\
\hline
30~GeV & 12 & 46.4 & 21.7 & 14.0 \\
50~GeV & 12 & 12.4 &  8.5 &  9.1 \\
70~GeV & 11 & 10.9 &  9.6 & 13.5 \\
90~GeV & 10 & 13.3 & 12.7 & 14.4
\end{tabular}
\end{ruledtabular}
\end{table}

\begin{figure*}[hbt]
  \begin{center}
   \includegraphics[width=15.6cm]{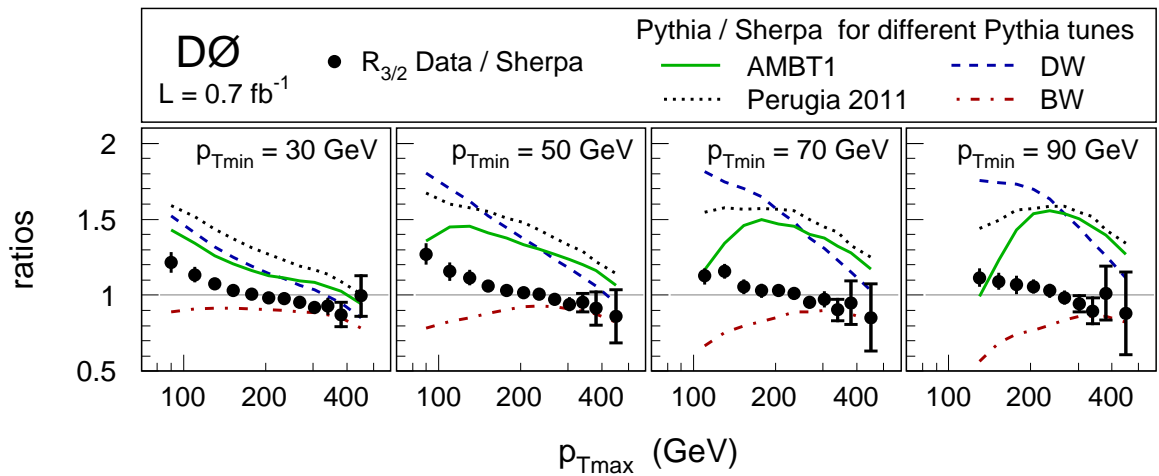}
  \caption{\label{fig:mccomp}
  The measured $\Rtt$ results, normalized to the predictions of the \sherpa\ Monte Carlo event generator. The inner uncertainty bars represent the statistical uncertainties
  while the total uncertainty bars represent the quadratic sums of statistical
  and systematic uncertainties. Overlaid are the predictions from the \pythia\ Monte Carlo event generator for four different tunes, also normalized to the \sherpa\ predictions.
  }
  \end{center}
\end{figure*}

The theoretical predictions for the MSTW2008NLO PDF sets are overlaid on
the data in Fig.~\ref{fig:result}.
The results for CT10 and NNPDFv2.1 PDFs (not shown) agree with those obtained for
MSTW2008NLO to better than 0.1\% for $\ptmax < 300$~GeV,
and are always better than 0.4\%.
Figure~\ref{fig:result} shows good agreement between the theoretical predictions and data.
At the lowest $\ptmin$ value, the ratio of data and theory decreases from $+8$\% to $-6$\% with increasing $\ptmax$.
For $\ptmin = 50$ and 70 GeV, the ratio of data to the theoretical predictions is consistent with
unity over the entire range of $\ptmax$. For $\ptmin = 90$~GeV the theoretical predictions are slightly higher
than the experimental results, but still consistent within the
theoretical uncertainty.
The agreement between theory and data is quantified
by computing $\chi^2$\,values for each choice of $\ptmin$. The $\chi^2$ definition takes into account all experimental uncertainties
and their correlations as well as uncertainties in the non-perturbative corrections and the PDFs.
The $\chi^2$ values are listed
in Table~\ref{tab:r32chi2} for the different $\ptmin$ requirements
and for three choices of $\mu_R$ and $\mu_F$.
For $\ptmin = 30$~GeV, the $\chi^2$ value depends strongly on the scale,
and agreement within the expectation of $\chi^2 = N_{\rm data} \pm \sqrt{2\cdot N_{\rm data}}$
(where $N_{\rm data}$ is the number of data points)
is obtained only for the largest scale of $\mu_{R,F} = 2\ptmax$.
For larger requirements of $\ptmin = 50$, $70$, and $90$~GeV,
the theoretical predictions agree with the data for all three scales,
and the central scale choice of $\mu_{R,F} = \ptmax$ always provides
the lowest $\chi^2$.

% ---------------------------------------------------------
% -----------     comparison with PYTHIA & SHERPA
% ---------------------------------------------------------
Predictions from different Monte Carlo event generators are compared
to the data in Fig.~\ref{fig:mccomp}.
In this Figure, the measured $\Rtt$\ results and the \pythia\ predictions
for different tunes are divided by the predictions from \sherpa\
which includes the tree-level matrix elements for
two-, three-, and four-jet production, matched with a parton shower.
The \sherpa\ predictions for $\Rtt$ have a different $\ptmax$
dependence and, independent of $\ptmin$, they are approximately 20\% lower (10\% higher)
than the data at low (high) $\ptmax$.

\pythia\ includes only the two-jet matrix elements and a parton shower.
The \pythia\ predictions for the three-jet cross section
therefore depend directly on the parton shower model
and the corresponding parameter settings (tunes).
The \pythia\ results have been obtained for
tunes DW, BW~\cite{Albrow:2006rt}, A~\cite{TuneA}, AMBT1, S Global~\cite{PerugiaS}, Perugia 2011, Perugia 2011 LO$^{**}$, and Perugia 2011 Tevatron~\cite{Perugia2011}.
The first three tunes use $Q^2$-ordered parton showers and an older underlying event model,
while the latter five  use $p_T$-ordered parton showers and a newer underlying event model.
Tune DW was tuned to describe the D0 measurement
of dijet azimuthal decorrelations~\cite{dphi},
and tunes AMBT1, S Global and Perugia 2011
were tuned to LHC data at $\sqrt{s}=7\,$TeV.
The predictions for tune Perugia 2011 agree within 1\% with those
for tunes Perugia 2011 LO$^{**}$ and Perugia 2011 Tevatron (the latter two
are not shown in Fig.~\ref{fig:mccomp}).
The predictions for tune A (not shown in Fig.~\ref{fig:mccomp}) are always
above those for tune DW,
and the predictions for tune S Global (not shown Fig.~\ref{fig:mccomp})
are 2--5\% higher than those for tune AMBT1.
Figure~\ref{fig:mccomp} shows that none of the studied \pythia\ tunes describe the data;
all predict a different $\ptmax$ dependence,
and the discrepancies are strongly depending on $\ptmin$.
While a dedicated study of the sensitivity of the \pythia\
parameters is beyond the scope of this letter,
these $\Rtt$\ data demonstrate the limitations of current
\pythia\ tunes and provide strong constraints for
future parameter adjustments.
% --------------------------------------------------------------
% -----------  Summary
% --------------------------------------------------------------

In summary, we have presented the first measurement of the ratio
$\Rtt$ of three-jet to two-jet cross sections in hadron-hadron collisions
at a center of mass energy of $\sqrt{s}=1.96\,$TeV.
The ratio $\Rtt$ is presented for $\ptmin$ requirements of 30, 50, 70, and 90 GeV,
as a function of the highest jet $p_T$, $\ptmax$, in the range of 80--500~GeV.
\sherpa\ predicts a slightly different $\ptmax$\ dependence, but it describes
the data within approximately $-10$\% to $+20$\%.
None of the \pythia\ tunes
DW, BW, A, AMBT1, S Global, and Perugia 2011
describe the data.
The data are well described by the pQCD predictions at the next-to-leading order
in the strong coupling constant $\as$, corrected for non-perturbative effects.

\begin{acknowledgments}
\input{acknowledgement.tex}
\end{acknowledgments}

\end{document}

%% file: author_list.tex
\affiliation{LAFEX, Centro Brasileiro de Pesquisas F\'{i}sicas, Rio de Janeiro, Brazil}
\affiliation{Universidade do Estado do Rio de Janeiro, Rio de Janeiro, Brazil}
\affiliation{Universidade Federal do ABC, Santo Andr\'e, Brazil}
\affiliation{University of Science and Technology of China, Hefei, People's Republic of China}
\affiliation{Universidad de los Andes, Bogot\'a, Colombia}
\affiliation{Charles University, Faculty of Mathematics and Physics, Center for Particle Physics, Prague, Czech Republic}
\affiliation{Czech Technical University in Prague, Prague, Czech Republic}
\affiliation{Center for Particle Physics, Institute of Physics, Academy of Sciences of the Czech Republic, Prague, Czech Republic}
\affiliation{Universidad San Francisco de Quito, Quito, Ecuador}
\affiliation{LPC, Universit\'e Blaise Pascal, CNRS/IN2P3, Clermont, France}
\affiliation{LPSC, Universit\'e Joseph Fourier Grenoble 1, CNRS/IN2P3, Institut National Polytechnique de Grenoble, Grenoble, France}
\affiliation{CPPM, Aix-Marseille Universit\'e, CNRS/IN2P3, Marseille, France}
\affiliation{LAL, Universit\'e Paris-Sud, CNRS/IN2P3, Orsay, France}
\affiliation{LPNHE, Universit\'es Paris VI and VII, CNRS/IN2P3, Paris, France}
\affiliation{CEA, Irfu, SPP, Saclay, France}
\affiliation{IPHC, Universit\'e de Strasbourg, CNRS/IN2P3, Strasbourg, France}
\affiliation{IPNL, Universit\'e Lyon 1, CNRS/IN2P3, Villeurbanne, France and Universit\'e de Lyon, Lyon, France}
\affiliation{III. Physikalisches Institut A, RWTH Aachen University, Aachen, Germany}
\affiliation{Physikalisches Institut, Universit\"at Freiburg, Freiburg, Germany}
\affiliation{II. Physikalisches Institut, Georg-August-Universit\"at G\"ottingen, G\"ottingen, Germany}
\affiliation{Institut f\"ur Physik, Universit\"at Mainz, Mainz, Germany}
\affiliation{Ludwig-Maximilians-Universit\"at M\"unchen, M\"unchen, Germany}
\affiliation{Fachbereich Physik, Bergische Universit\"at Wuppertal, Wuppertal, Germany}
\affiliation{Panjab University, Chandigarh, India}
\affiliation{Delhi University, Delhi, India}
\affiliation{Tata Institute of Fundamental Research, Mumbai, India}
\affiliation{University College Dublin, Dublin, Ireland}
\affiliation{Korea Detector Laboratory, Korea University, Seoul, Korea}
\affiliation{CINVESTAV, Mexico City, Mexico}
\affiliation{Nikhef, Science Park, Amsterdam, the Netherlands}
\affiliation{Radboud University Nijmegen, Nijmegen, the Netherlands}
\affiliation{Joint Institute for Nuclear Research, Dubna, Russia}
\affiliation{Institute for Theoretical and Experimental Physics, Moscow, Russia}
\affiliation{Moscow State University, Moscow, Russia}
\affiliation{Institute for High Energy Physics, Protvino, Russia}
\affiliation{Petersburg Nuclear Physics Institute, St. Petersburg, Russia}
\affiliation{Instituci\'{o} Catalana de Recerca i Estudis Avan\c{c}ats (ICREA) and Institut de F\'{i}sica d'Altes Energies (IFAE), Barcelona, Spain}
\affiliation{Uppsala University, Uppsala, Sweden}
\affiliation{Lancaster University, Lancaster LA1 4YB, United Kingdom}
\affiliation{Imperial College London, London SW7 2AZ, United Kingdom}
\affiliation{The University of Manchester, Manchester M13 9PL, United Kingdom}
\affiliation{University of Arizona, Tucson, Arizona 85721, USA}
\affiliation{University of California Riverside, Riverside, California 92521, USA}
\affiliation{Florida State University, Tallahassee, Florida 32306, USA}
\affiliation{Fermi National Accelerator Laboratory, Batavia, Illinois 60510, USA}
\affiliation{University of Illinois at Chicago, Chicago, Illinois 60607, USA}
\affiliation{Northern Illinois University, DeKalb, Illinois 60115, USA}
\affiliation{Northwestern University, Evanston, Illinois 60208, USA}
\affiliation{Indiana University, Bloomington, Indiana 47405, USA}
\affiliation{Purdue University Calumet, Hammond, Indiana 46323, USA}
\affiliation{University of Notre Dame, Notre Dame, Indiana 46556, USA}
\affiliation{Iowa State University, Ames, Iowa 50011, USA}
\affiliation{University of Kansas, Lawrence, Kansas 66045, USA}
\affiliation{Kansas State University, Manhattan, Kansas 66506, USA}
\affiliation{Louisiana Tech University, Ruston, Louisiana 71272, USA}
\affiliation{Northeastern University, Boston, Massachusetts 02115, USA}
\affiliation{University of Michigan, Ann Arbor, Michigan 48109, USA}
\affiliation{Michigan State University, East Lansing, Michigan 48824, USA}
\affiliation{University of Mississippi, University, Mississippi 38677, USA}
\affiliation{University of Nebraska, Lincoln, Nebraska 68588, USA}
\affiliation{Rutgers University, Piscataway, New Jersey 08855, USA}
\affiliation{Princeton University, Princeton, New Jersey 08544, USA}
\affiliation{State University of New York, Buffalo, New York 14260, USA}
\affiliation{University of Rochester, Rochester, New York 14627, USA}
\affiliation{State University of New York, Stony Brook, New York 11794, USA}
\affiliation{Brookhaven National Laboratory, Upton, New York 11973, USA}
\affiliation{Langston University, Langston, Oklahoma 73050, USA}
\affiliation{University of Oklahoma, Norman, Oklahoma 73019, USA}
\affiliation{Oklahoma State University, Stillwater, Oklahoma 74078, USA}
\affiliation{Brown University, Providence, Rhode Island 02912, USA}
\affiliation{University of Texas, Arlington, Texas 76019, USA}
\affiliation{Southern Methodist University, Dallas, Texas 75275, USA}
\affiliation{Rice University, Houston, Texas 77005, USA}
\affiliation{University of Virginia, Charlottesville, Virginia 22904, USA}
\affiliation{University of Washington, Seattle, Washington 98195, USA}
\author{V.M.~Abazov} \affiliation{Joint Institute for Nuclear Research, Dubna, Russia}
\author{B.~Abbott} \affiliation{University of Oklahoma, Norman, Oklahoma 73019, USA}
\author{B.S.~Acharya} \affiliation{Tata Institute of Fundamental Research, Mumbai, India}
\author{M.~Adams} \affiliation{University of Illinois at Chicago, Chicago, Illinois 60607, USA}
\author{T.~Adams} \affiliation{Florida State University, Tallahassee, Florida 32306, USA}
\author{G.D.~Alexeev} \affiliation{Joint Institute for Nuclear Research, Dubna, Russia}
\author{G.~Alkhazov} \affiliation{Petersburg Nuclear Physics Institute, St. Petersburg, Russia}
\author{A.~Alton$^{a}$} \affiliation{University of Michigan, Ann Arbor, Michigan 48109, USA}
\author{A.~Askew} \affiliation{Florida State University, Tallahassee, Florida 32306, USA}
\author{S.~Atkins} \affiliation{Louisiana Tech University, Ruston, Louisiana 71272, USA}
% ...
%
% visitor_addresses.tex                       8 August 2012
%  available symbols are:
%  $\ast, \dag, \ddag, \S, \P, $\|$, $\ast\ast$, \dag\dag, \ddag\ddag ,\#
%
\collaboration{The D0 Collaboration\footnote{with visitors from
%{alton}
$^{a}$Augustana College, Sioux Falls, SD, USA,
%{burdin}
$^{b}$The University of Liverpool, Liverpool, UK,
%{garcia-guerra}
$^{c}$UPIITA-IPN, Mexico City, Mexico,
%{grohsjean}
$^{d}$DESY, Hamburg, Germany,
%{partridge}
$^{e}$SLAC, Menlo Park, CA, USA,
%{hesketh}
$^{f}$University College London, London, UK,
%{luna-garcia}
$^{g}$Centro de Investigacion en Computacion - IPN, Mexico City, Mexico,
%{podesta-lerma}
$^{h}$ECFM, Universidad Autonoma de Sinaloa, Culiac\'an, Mexico
and
%{santos}
$^{i}$Universidade Estadual Paulista, S\~ao Paulo, Brazil.
%{falkowski}
%$^{?}$Laboratoire de Physique Theorique, Orsay, FR,
%{hooper}
%$^{?}$Visitor from Bradley University, Peoria, IL, USA.
%{kozminski}
%$^{?}$}Visitor from Lewis University, Romeoville, IL, USA.
%{weber}
%$^{?}$Universit{\"a}t Bern, Bern, Switzerland.
%{deceased}
%$^{\ddag}$Deceased.
}} \noaffiliation
\vskip 0.25cm

%% file: R32table.tex
\begin{table}[h]
\centering
\caption{\label{tab:consistency}
The ratio $\Rtt$\ measured as a function of $\ptmax$ for different
$\ptmin$\ requirements, along with statistical and systematic uncertainties.
}
\begin{ruledtabular}
\begin{tabular}{rcrccc}
 $p_{T{\rm max}}$  &  $p_{T{\rm min}}$ & \multicolumn{1}{c}{$R_{3/2}$} & Stat. uncert. & \multicolumn{2}{c}{Syst. uncert.} \\
 (GeV) &  (GeV) & & (percent) & \multicolumn{2}{c}{(percent)} \\
\hline
 $ 80$--$100$  &  $ 30$  &  $ 1.816 \times 10^{-1}$  & \phantom{00}  $\pm  0.7$  &  $+  5.6$  &  $-  5.5$ \\
 $100$--$120$  &  $ 30$  &  $ 2.182 \times 10^{-1}$  & \phantom{00} $\pm  0.6$  &  $+  4.5$  &  $-  4.4$ \\
 $120$--$140$  &  $ 30$  &  $ 2.370 \times 10^{-1}$  & \phantom{00} $\pm  0.5$  &  $+  3.7$  &  $-  3.7$ \\
 $140$--$165$  &  $ 30$  &  $ 2.442 \times 10^{-1}$  & \phantom{00} $\pm  0.6$  &  $+  3.3$  &  $-  3.3$ \\
 $165$--$190$  &  $ 30$  &  $ 2.464 \times 10^{-1}$  & \phantom{00} $\pm  1.0$  &  $+  3.1$  &  $-  3.2$ \\
 $190$--$220$  &  $ 30$  &  $ 2.421 \times 10^{-1}$  & \phantom{00} $\pm  0.6$  &  $+  3.1$  &  $-  3.1$ \\
 $220$--$250$  &  $ 30$  &  $ 2.362 \times 10^{-1}$  & \phantom{00} $\pm  0.9$  &  $+  3.1$  &  $-  3.1$ \\
 $250$--$285$  &  $ 30$  &  $ 2.228 \times 10^{-1}$  & \phantom{00} $\pm  1.4$  &  $+  3.3$  &  $-  3.2$ \\
 $285$--$320$  &  $ 30$  &  $ 2.021 \times 10^{-1}$  & \phantom{00} $\pm  2.7$  &  $+  3.5$  &  $-  3.4$ \\
 $320$--$360$  &  $ 30$  &  $ 1.925 \times 10^{-1}$  & \phantom{00} $\pm  4.6$  &  $+  3.8$  &  $-  3.8$ \\
 $360$--$400$  &  $ 30$  &  $ 1.688 \times 10^{-1}$  & \phantom{00} $\pm  9.1$  &  $+  4.1$  &  $-  4.2$ \\
 $400$--$500$  &  $ 30$  &  $ 1.814 \times 10^{-1}$  & \phantom{0} $\pm 13.4$  &  $+  4.6$  &  $-  4.6$ \\
 $ 80$--$100$  &  $ 50$  &  $ 3.116 \times 10^{-2}$  & \phantom{00} $\pm  1.5$  &  $+  5.5$  &  $-  5.5$ \\
 $100$--$120$  &  $ 50$  &  $ 6.796 \times 10^{-2}$  & \phantom{00} $\pm  1.6$  &  $+  5.1$  &  $-  5.1$ \\
 $120$--$140$  &  $ 50$  &  $ 1.059 \times 10^{-1}$  & \phantom{00} $\pm  1.4$  &  $+  4.5$  &  $-  4.6$ \\
 $140$--$165$  &  $ 50$  &  $ 1.292 \times 10^{-1}$  & \phantom{00} $\pm  1.0$  &  $+  3.5$  &  $-  3.5$ \\
 $165$--$190$  &  $ 50$  &  $ 1.420 \times 10^{-1}$  & \phantom{00} $\pm  1.4$  &  $+  3.1$  &  $-  3.2$ \\
 $190$--$220$  &  $ 50$  &  $ 1.477 \times 10^{-1}$  & \phantom{00} $\pm  0.8$  &  $+  2.8$  &  $-  2.9$ \\
 $220$--$250$  &  $ 50$  &  $ 1.470 \times 10^{-1}$  & \phantom{00} $\pm  1.2$  &  $+  2.8$  &  $-  2.8$ \\
 $250$--$285$  &  $ 50$  &  $ 1.398 \times 10^{-1}$  & \phantom{00} $\pm  1.9$  &  $+  3.0$  &  $-  2.8$ \\
 $285$--$320$  &  $ 50$  &  $ 1.290 \times 10^{-1}$  & \phantom{00} $\pm  3.6$  &  $+  3.3$  &  $-  3.0$ \\
 $320$--$360$  &  $ 50$  &  $ 1.217 \times 10^{-1}$  & \phantom{00} $\pm  6.2$  &  $+  3.5$  &  $-  3.4$ \\
 $360$--$400$  &  $ 50$  &  $ 1.071 \times 10^{-1}$  & \phantom{0} $\pm 12.2$  &  $+  3.8$  &  $-  3.9$ \\
 $400$--$500$  &  $ 50$  &  $ 9.105 \times 10^{-2}$  & \phantom{0} $\pm 20.4$  &  $+  4.5$  &  $-  4.4$ \\
 $100$--$120$  &  $ 70$  &  $ 1.161 \times 10^{-2}$  & \phantom{00} $\pm  2.2$  &  $+  4.1$  &  $-  4.4$ \\
 $120$--$140$  &  $ 70$  &  $ 2.699 \times 10^{-2}$  & \phantom{00} $\pm  1.3$  &  $+  4.1$  &  $-  4.2$ \\
 $140$--$165$  &  $ 70$  &  $ 4.849 \times 10^{-2}$  & \phantom{00} $\pm  1.7$  &  $+  4.3$  &  $-  4.3$ \\
 $165$--$190$  &  $ 70$  &  $ 7.254 \times 10^{-2}$  & \phantom{00} $\pm  2.2$  &  $+  4.0$  &  $-  4.1$ \\
 $190$--$220$  &  $ 70$  &  $ 8.880 \times 10^{-2}$  & \phantom{00} $\pm  1.1$  &  $+  3.3$  &  $-  3.5$ \\
 $220$--$250$  &  $ 70$  &  $ 9.401 \times 10^{-2}$  & \phantom{00} $\pm  1.6$  &  $+  3.2$  &  $-  3.3$ \\
 $250$--$285$  &  $ 70$  &  $ 9.125 \times 10^{-2}$  & \phantom{00} $\pm  2.5$  &  $+  3.2$  &  $-  3.0$ \\
 $285$--$320$  &  $ 70$  &  $ 8.969 \times 10^{-2}$  & \phantom{00} $\pm  4.5$  &  $+  3.5$  &  $-  3.1$ \\
 $320$--$360$  &  $ 70$  &  $ 7.852 \times 10^{-2}$  & \phantom{00} $\pm  7.9$  &  $+  3.6$  &  $-  3.5$ \\
 $360$--$400$  &  $ 70$  &  $ 7.555 \times 10^{-2}$  & \phantom{0} $\pm 14.9$  &  $+  4.0$  &  $-  4.1$ \\
 $400$--$500$  &  $ 70$  &  $ 5.959 \times 10^{-2}$  & \phantom{0} $\pm 26.0$  &  $+  5.0$  &  $-  4.8$ \\
 $120$--$140$  &  $ 90$  &  $ 5.775 \times 10^{-3}$  & \phantom{00} $\pm  2.9$  &  $+  4.7$  &  $-  4.7$ \\
 $140$--$165$  &  $ 90$  &  $ 1.281 \times 10^{-2}$  & \phantom{00} $\pm  2.8$  &  $+  4.3$  &  $-  4.4$ \\
 $165$--$190$  &  $ 90$  &  $ 2.564 \times 10^{-2}$  & \phantom{00} $\pm  3.5$  &  $+  4.6$  &  $-  4.6$ \\
 $190$--$220$  &  $ 90$  &  $ 4.435 \times 10^{-2}$  & \phantom{00} $\pm  1.7$  &  $+  4.2$  &  $-  4.2$ \\
 $220$--$250$  &  $ 90$  &  $ 5.744 \times 10^{-2}$  & \phantom{00} $\pm  2.2$  &  $+  3.9$  &  $-  3.9$ \\
 $250$--$285$  &  $ 90$  &  $ 6.122 \times 10^{-2}$  & \phantom{00} $\pm  3.2$  &  $+  3.6$  &  $-  3.4$ \\
 $285$--$320$  &  $ 90$  &  $ 6.002 \times 10^{-2}$  & \phantom{00} $\pm  5.6$  &  $+  3.6$  &  $-  3.4$ \\
 $320$--$360$  &  $ 90$  &  $ 5.482 \times 10^{-2}$  & \phantom{00} $\pm  9.7$  &  $+  3.8$  &  $-  3.8$ \\
 $360$--$400$  &  $ 90$  &  $ 5.685 \times 10^{-2}$  & \phantom{0} $\pm 17.5$  &  $+  4.2$  &  $-  4.3$ \\
 $400$--$500$  &  $ 90$  &  $ 4.327 \times 10^{-2}$  & \phantom{0} $\pm 31.0$  &  $+  5.4$  &  $-  4.7$ 
\end{tabular}
\end{ruledtabular}
\end{table}

%% file: acknowledgement.tex
% acknowledgement.tex                            8 August 2012 
%
We thank the staffs at Fermilab and collaborating institutions,
and acknowledge support from the
DOE and NSF (USA);
CEA and CNRS/IN2P3 (France);
MON, NRC KI and RFBR (Russia);
CNPq, FAPERJ, FAPESP and FUNDUNESP (Brazil);
DAE and DST (India);
Colciencias (Colombia);
CONACyT (Mexico);
NRF (Korea);
FOM (The Netherlands);
STFC and the Royal Society (United Kingdom);
MSMT and GACR (Czech Republic);
BMBF and DFG (Germany);
SFI (Ireland);
The Swedish Research Council (Sweden);
and
CAS and CNSF (China).